\documentclass[conference]{IEEEtran}

\usepackage{amsthm}
\usepackage{times}
\usepackage{breqn}
\usepackage{algorithm}
\usepackage{algpseudocode}
\algrenewcommand\algorithmicrequire{\textbf{Input:}}
\algrenewcommand\algorithmicensure{\textbf{Output:}}

\usepackage{epsfig}
\usepackage{epstopdf}
\usepackage{subfigure}
\usepackage{multirow}
\usepackage{comment}
\usepackage{listings}
\usepackage{xcolor}
\usepackage{soul}
\usepackage{xspace}
\usepackage{url}
\usepackage{rotating}

\newcommand{\bigO}[1]{$\mathcal{O}\paren{#1}$\xspace}
\newcommand{\bigOmega}[1]{\mbox{\mbox{$\Omega$}$\paren{#1}$}\xspace}
\newcommand{\bigTheta}[1]{\mbox{\mbox{$\Theta$}$\paren{#1}$}\xspace}
\newcommand{\paren}[1]{\left(  #1 \right)}
\newcommand{\edge}[1]{(  #1 )}

\newtheorem{theorem}{Theorem}

\newtheorem{lemma}{Lemma}

\newtheorem{definition}[theorem]{Definition}

\usepackage{cite}
\usepackage{amsmath,amssymb,amsfonts}
\usepackage{graphicx}
\usepackage{textcomp}
\usepackage{xcolor}
\def\BibTeX{{\rm B\kern-.05em{\sc i\kern-.025em b}\kern-.08em
    T\kern-.1667em\lower.7ex\hbox{E}\kern-.125emX}}

\graphicspath{{figures/}}

\makeatletter
\newcommand{\linebreakand}{%
  \end{@IEEEauthorhalign}
  \hfill\mbox{}\par
  \mbox{}\hfill\begin{@IEEEauthorhalign}
}
\makeatother

\begin{document}

\title{Fast Triangle Counting}
\author{

\IEEEauthorblockN{David A. Bader$^*$}
\IEEEauthorblockA{\textit{Department of Data Science} \\
\textit{New Jersey Institute of Technology}\\
Newark, New Jersey, USA\\
bader@njit.edu}

}

\maketitle

\begingroup\renewcommand\thefootnote{$^*$}
\footnotetext{This research was partially supported by NSF grant number CCF-2109988.}
\endgroup
\begin{abstract}
Listing and counting triangles in graphs is a key algorithmic kernel for network analyses including community detection, clustering coefficients, k-trusses, and triangle centrality. We design and implement a new serial algorithm for triangle counting that performs competitively with the fastest previous approaches on both real and synthetic graphs, such as those from the Graph500 Benchmark and the MIT/Amazon/IEEE Graph Challenge. The experimental results use the recently-launched Intel Xeon Platinum 8480+ and CPU Max 9480 processors.
\end{abstract}


\section{Introduction}

Triangle listing and counting is a highly-studied problem in computer science and is a key building block in various graph analysis techniques such as clustering coefficients \cite{watts1998collective}, k-truss \cite{cohen2008trusses}, and triangle centrality \cite{burkhardt2021triangle}. The MIT/Amazon/IEEE Graph Challenge \cite{Samsi2018,Samsi2020} includes triangle counting as a fundamental method in graph analytics.
There are at most ${n \choose 3} = \bigTheta{n^3}$ triangles in a graph $G = (V, E)$ with $n = |V|$ vertices and $m=|E|$ edges.
The focus of this paper is on sequential triangle counting algorithms for sparse graphs that are stored in compressed, sparse row (CSR) format, rather than adjacency matrix format.
The na\"ive approach using triply-nest loops to check if each triple $(u, v, w)$ forms a triangle takes \bigO{n^3} time and is inefficient for sparse graphs.  It is well-known that listing all triangles in G is $\bigOmega{m^{\frac{3}{2}}}$ time \cite{itai1978finding, latapy2007practical}.

The main contributions of this paper are:
\begin{itemize}
    \item A new triangle algorithm that combines the techniques of cover-edges, forward, and hashing and runs in \bigO{m \cdot a(G)}, where $a(G)$ is the arboricity of the graph; 
    \item An experimental study of an implementation of this novel triangle counting algorithm on real and synthetic graphs; and
    \item Freely-available, open-source software for more than 20 triangle counting algorithms and variants in the C programming language.
\end{itemize}

\subsection{Related work}

There are faster algorithms for triangle counting, such as the work of Alon, Yuster, and Zwick \cite{alon1997finding} that require an adjacency matrix for the input graph representation and use fast matrix multiplication. As this is infeasible for large, sparse graph, their and other fast multiply methods are outside the scope of this paper. 

Latapy \cite{latapy2007practical} provides a survey on triangle counting algorithms for very large, sparse graphs. One of the earliest algorithms, \emph{tree-listing}, published in 1978 by Itai and Rodeh \cite{itai1978finding} first finds a rooted spanning tree of the graph. After iterating through the non-tree edges and using criteria to identify triangles, the tree edges are removed and the algorithm repeats until there are no edges remaining. This approach takes \bigO{m^{\frac{3}{2}}} time (or \bigO{n} for planar graphs).

The most common triangle counting algorithms in the literature include \emph{vertex-iterator} \cite{itai1978finding,latapy2007practical} and \emph{edge-iterator} \cite{itai1978finding,latapy2007practical} approaches that run in 
\bigO{m \cdot d_{\mbox{max}}} time \cite{itai1978finding, schank2005finding, schank2007}, where $d_{\mbox{max}}$ is the maximum degree of a vertex in the graph. 
In vertex-iterator, the adjacency list $N(v)$ of each vertex $v \in V$ is doubly-enumerated to find all 2-paths $(u, v, w)$ where $u, w \in N(v)$. Then, the graph is searched for the existence of the closing edge $(u, w)$ by checking if $w \in N(u)$ (or if $u \in N(w)$).  Arifuzzaman~\emph{et al.} \cite{arifuzzaman2019} study modifications of the vertex-iterator algorithm based on various methods for vertex ordering.

In edge-iterator, each edge $(u, v)$ in the graph is examined, and the intersection of $N(u)$ and $N(v)$ is computed to find triangles. A common optimization is to use a \emph{direction-oriented} approach that only considers edges $(u, v)$ where $u< v$.  The variants of edge-iterator are often based on the algorithm used to perform the intersection. When the two adjacency lists are sorted, then \emph{MergePath} and \emph{BinarySearch} can be used. MergePath performs a linear scan through both lists counting the common elements. Makkar, Bader and Green \cite{makkar2017} give an efficient MergePath algorithm for GPU. Mailthody~\emph{et al.} \cite{mailthody2018} use an optimized two-pointer intersection (MergePath) for set intersection.
BinarySearch, as the name implies, uses a binary search to determine if each element of the smaller list is found in the larger list. \emph{Hash} is another method for performing the intersection of two sets and it does not require the adjacency lists to be sorted. A typical implementation of Hash initializes a Boolean array of size $m$ to all false. Then, positions in Hash corresponding to the vertex values in $N(u)$ are set to true. Then $N(v)$ is scanned, looking up in $\bigTheta{1}$ time whether or not there is a match for each vertex. Chiba and Nishizeki published one of the earliest edge iterator with hashing algorithms for triangle finding in 1985 \cite{chiba1985}. The running time is \bigO{a(G) m}, where $a(G)$ is defined as the arboricity of $G$, which is upper-bounded $a(G) \leq \lceil (2m + n) ^{\frac{1}{2}} / 2 \rceil$ \cite{chiba1985}. In 2018, Davis rediscovered this method, which he calls \texttt{tri\_simple} in his comparison with SuiteSparse GraphBLAS \cite{davis2018HPEC}. According to Davis \cite{davis2018HPEC}: this algorithm ``is already a non-trivial method. It requires expert knowledge of how Gustavson’s
method can be implemented efficiently, including a reduction of the result to a single scalar.'' Mowlaei \cite{Mowlaei2017} gave a variant of the edge-iterator algorithm that uses vectorized sorted set intersection and reorders the vertices using the reverse Cuthill-McKee heuristic.

In 2005, Schank and Wagner \cite{schank2005finding,schank2007} designed a fast triangle counting algorithm called \emph{forward} (see Algorithm~\ref{alg:forward}) that is a refinement of the edge-iterator approach. Instead of intersections of the full adjacency lists, the \emph{forward} algorithm uses a dynamic data structure $A(v)$ to store a subset of the neighborhood $N(v)$ for $v \in V$. Initially each set $A()$ is empty, and after computing the intersection of the sets $A(u)$ and $A(v)$ for each edge $(u, v)$ (with $u<v$), $u$ is added to $A(v)$. This significantly reduces the size of the intersections needed to find triangles. The running time is \bigO{m \cdot d_{\mbox{max}}}. However, if one reorders the vertices in decreasing order of their degrees as a \bigTheta{n \log n} time pre-processing step, the forward algorithm's running time reduces to \bigO{m^{\frac{3}{2}}}. Donato~\emph{et al.} \cite{donato2018} implement the forward algorithm for shared-memory. Ortmann and Brandes \cite{ortmann2014} survey triangle counting algorithms, create a unifying framework for parsimonious implementations, and conclude that nearly every triangle listing variant is in \bigO{m \cdot a(G)}. 

\begin{algorithm}[htbp]
\footnotesize
\caption{Forward Triangle Counting \cite{schank2005finding,schank2007}}
\label{alg:forward}
\begin{algorithmic}[1]
\Require{Graph $G = (V, E)$}
\Ensure{Triangle Count $T$}
\State $T \leftarrow 0$
\State $\forall v \in V$
    \State \hspace{8pt} $A(v) \leftarrow \emptyset$
\State $\forall \edge{u, v} \in E$
    \State \hspace{8pt} if $(u<v)$ then
        \State \hspace{16pt} $\forall w \in A(u) \cap A(v)$
            \State \hspace{24pt} $T \leftarrow T+1$
        \State \hspace{16pt} $A(v) \leftarrow A(v) \cup \{ u\}$
\end{algorithmic}
\end{algorithm}

The \emph{forward-hashed} algorithm \cite{schank2005finding,schank2007} (also called \emph{compact-forward} \cite{latapy2007practical}) is a variant of the forward algorithm that uses the hashing described above for the intersections of the $A()$ sets, see Algorithm~\ref{alg:forwardhash}.  Shun and Tangwongsan \cite{shun2015multicore} parallelize the forward and forward-hashed algorithms for multicore systems. Low~\emph{et al.} \cite{Low2017} derive a linear-algebra method for triangle counting that does not use matrix multiplication. Their algorithm results in the forward-hashed algorithm. 

\begin{algorithm}[htbp]
\footnotesize
\caption{Forward-Hashed Triangle Counting \cite{schank2005finding,schank2007}}
\label{alg:forwardhash}
\begin{algorithmic}[1]
\Require{Graph $G = (V, E)$}
\Ensure{Triangle Count $T$}
\State $T \leftarrow 0$
\State $\forall v \in V$
    \State \hspace{8pt} $A(v) \leftarrow \emptyset$
\State $\forall \edge{u, v} \in E$
    \State \hspace{8pt} if $(u<v)$ then
        \State \hspace{16pt} $\forall w \in A(u)$
            \State \hspace{24pt} Hash[$w$] $\leftarrow$ true
        \State \hspace{16pt} $\forall w \in A(v)$
            \State \hspace{24pt} if Hash[$w$] then
                \State \hspace{32pt} $T \leftarrow T+1$
        \State \hspace{16pt} $\forall w \in A(u)$
            \State \hspace{24pt} Hash[$w$] $\leftarrow$ false
        \State \hspace{16pt} $A(v) \leftarrow A(v) \cup \{ u\}$
\end{algorithmic}
\end{algorithm}

\section{Algorithm}

Recently, we presented Algorithm~\ref{alg:CETC} \cite{2023-BLGGLRD} as a new method for finding triangles. This approach finds a subset of \emph{cover edges} from $E$ such that every triangle contains at least one cover edge. 

\begin{algorithm}[htbp]
\footnotesize
\caption{Cover-Edge Triangle Counting \cite{2023-BLGGLRD}}
\label{alg:CETC}
\begin{algorithmic}[1]
\Require{Graph $G = (V, E)$}
\Ensure{Triangle Count $T$}
\State $T \leftarrow 0$  \label{l:init}
\State $\forall v \in V$ \label{l:bfs1}
\State \hspace{8pt} if $v$ unvisited, then BFS($G$, $v$) \label{l:bfs2}
\State $\forall \edge{u, v} \in E$ \label{l:edge1}
\State \hspace{8pt} if $(L(u) \equiv L(v)) \land (u < v)$ \label{l:horiz} \Comment{$\edge{u,v}$ is horizontal}
\State \hspace{16pt} $\forall w \in N(u) \cap N(v)$ \label{l:intersect}
\State \hspace{24pt} if $(L(u)\neq L(w)) \lor \left( (L(u) \equiv L(w)) \land (v<w) \right)$ then \label{l:logic}
\State \hspace{32pt}  $T\leftarrow T + 1 $     \label{l:edge2} 
\end{algorithmic}
\end{algorithm}

This algorithm uses breadth-first search (BFS) to find a reduced cover-edge set consisting of edges $(u, v)$ where the levels of vertices $u$ and $v$ are the same, i.e., $L(u) \equiv L(v)$. 
From the result in \cite{2023-BLGGLRD}, each triangle must contain at least one of these horizontal edges.
Then each edge in the cover set is examined, and Hash is used to find the vertices $w$ in the intersection of $N(u)$ and $N(v)$. A triangle $(u, v, w)$ is found based on logic about $w$'s level. The breadth-first search, including determining the level of each vertex and marking horizontal-edges, requires \bigO{n+m} time. The number of horizontal edges is \bigO{m}. 
The intersection of each pair of vertices costs \bigO{d_{\mbox{max}}}.  
Hence, Alg.~\ref{alg:CETC} has complexity \bigO{m \cdot d_{\text{max}}}.

\begin{algorithm}[htbp]
\footnotesize
\caption{Fast Triangle Counting }
\label{alg:TC}
\begin{algorithmic}[1]
\Require{Graph $G = (V, E)$}
\Ensure{Triangle Count $T$}
\State $\forall v \in V$ \label{l:tc:bfs1}
    \State \hspace{8pt} if $v$ unvisited, then BFS($G$, $v$) \label{l:tc:bfs2}
\State $\forall \edge{u, v} \in E$ \label{l:tc:edge1}
    \State \hspace{8pt} if $(L(u) \equiv L(v))$ then \Comment{$\edge{u,v}$ is horizontal}
        \State \hspace{16pt} Add $(u,v)$ to $G0$
    \State \hspace{8pt} else
        \State \hspace{16pt} Add $(u,v)$ to $G1$ \label{l:tc:edge2}
\State $T \leftarrow \mbox{TC}\_{\mbox{forward-hashed}}(G0)$ \Comment{Alg.~\ref{alg:forwardhash}} \label{l:tc:forwardhashed}
\State $\forall u \in V_{G1}$ \label{l:tc:hash1}
    \State \hspace{8pt} $\forall v \in N_{G1}(u)$
        \State \hspace{16pt} Hash[$v$] $\leftarrow$ true
    \State \hspace{8pt} $\forall v \in N_{G0}(u)$
        \State \hspace{16pt} if $(u < v)$ then
            \State \hspace{24pt} $\forall w \in N_{G1}(v)$
                \State \hspace{32pt} if Hash[$w$] then
                    \State \hspace{40pt} $T \leftarrow T+1$
    \State \hspace{8pt} $\forall v \in N_{G1}(u)$
        \State \hspace{16pt} Hash[$v$] $\leftarrow$ false \label{l:tc:hash2}
\end{algorithmic}
\end{algorithm}

In this paper, we present our new triangle counting algorithm (Alg.~\ref{alg:TC}), called \emph{fast triangle counting}. This new triangle counting algorithm is similar with cover-edge triangle counting in Alg.~\ref{alg:CETC} and uses BFS to assign a level to each vertex in lines \ref{l:tc:bfs1} and \ref{l:tc:bfs2}.
Next in  lines~\ref{l:tc:edge1} to \ref{l:tc:edge2}, the edges $E$ of the graph are partitioned into two sets $E0$ -- the horizontal edges where both endpoints are on the same level -- and $E1$ -- the remaining tree and non-tree edges that span a level. Thus, we now have two graphs, $G0 = (V, E0)$ and $G1 = (V, E1)$, where $E = E0 \cup E1$ and $E0 \cap E1 = \emptyset$.  Triangles that are fully in $G0$ are counted with one method and triangles not fully in $G0$ are counted with another method.  For $G0$, the graph with horizontal edges, we count the triangles efficiently using the forward-hashed method (line~\ref{l:tc:forwardhashed}). For triangles not fully in $G0$, the algorithm uses the following approach to count these triangles. Using $G1$, the graph that contains the edges that span levels, we use a hashed intersection approach in lines~\ref{l:tc:hash1} to \ref{l:tc:hash2}. As per the cover-edge triangle counting, we need to find the intersections of the adjacency lists from the endpoints of horizontal edges. Thus, we use $G0$ to select the edges, and perform the hash-based intersections from the adjacency lists in graph $G1$. The proof of correctness for cover-edge triangle counting is given in \cite{2023-BLGGLRD}. Alg.~\ref{alg:TC} is a hybrid version of this algorithm, that partitions the edge set, and uses two different methods to count these two types of triangles. The proof of correctness is still valid with these new refinements to the algorithm. The running time of Alg.~\ref{alg:TC} is the maximum of the running time of forward-hashing and Alg.~\ref{alg:CETC}. Alg.~\ref{alg:TC} uses hashing for the set intersections. For vertices $u$ and $v$ the cost is $\min(d(u), d(v))$ since the algorithm can check if the neighbors of the lower-degree endpoint are in the hash set of the higher-degree endpoint. Over all $(u,v)$ edges in $E$, these intersections take \bigO{m \cdot a(G)} expected time. Hence, Alg.~\ref{alg:TC} takes \bigO{m \cdot a(G)} expected time. 

Similar with the forward-hashed method, by pre-processing the graph by re-ordering the vertices in decreasing order of degree in \bigTheta{n \log n} time often leads to a faster triangle counting algorithm in practice.

\section{Experimental Results}

\looseness=-1
We implemented more than 20 triangle counting algorithms and variants in C and use the Intel Development Cloud for benchmarking our results on a GNU/Linux node. The compiler is Intel(R) oneAPI DPC++/C++ Compiler 2023.1.0 (2023.1.0.20230320) and `\texttt{-O2}` is used as a compiler optimization flag.  For benchmarking we compare the performance using two recently-launched Intel Xeon processors (Sapphire Rapids launched Q1'23) with two types of memory (DDR5 and HBM). The first node is a dedicated 2.00 GHz 56-core (112 thread) Intel(R) Xeon(R) Platinum 8480+ processor (formerly known as Sapphire Rapids) with 105M cache and 1024GB of DDR5 RAM. The second node is a dedicated 1.90 GHz 56-core (112 thread) Intel(R) Xeon(R) CPU Max 9480 processor (formerly known as Sapphire Rapids HBM) with 112.5M cache and 256GB of high-memory bandwidth (HBM) memory. 

Following the best practices of experimental algorithmics \cite{McGeoch2012}, we conduct the benchmarking as follows.
Each algorithm is written in C and has a single argument -- a pointer to the graph in a compressed sparse row (CSR) format. The input is treated as read-only. If the implementation needs auxiliary arrays, pre-processing steps, or additional data structures, it is charged the full cost. Each implementation must manage memory and not contain any memory leaks -- hence, any dynamically allocated memory must be freed prior to returning the result.  The output from each implementation is an integer with the number of triangles found.  Each algorithm is run ten times, and the mean running time is reported.  To reduce variance for random graphs, the same graph instance is used for all of the experiments.  The source code is sequential C code without any explicit parallelization. The same coding style and effort was used for each implementation. 

Experimental results are presented in Table~\ref{t:results:intel8480} for the Intel Xeon Platinum 8480+ processor with DDR5 memory and in Table~\ref{t:results:intel9480} for the Intel Xeon Max 9480 processor with HBM memory. For each graph, we give the number of vertices ($n$), the number of edges ($m$), the number of triangles, and $k$ -- the percentage of graph edges that are horizontal after running BFS from arbitrary roots. The algorithms tested are
\begin{description}
\item[IR]: Treelist from Itai-Rodeh \cite{itai1978finding}
\item[V]: Vertex-iterator
\item[VD]: Vertex Iterator (direction-oriented)
\item[EM]: Edge Iterator with MergePath for set intersection
\item[EMD]: Edge Iterator with MergePath for set intersection (direction-oriented)
\item[EB]: Edge Iterator with BinarySearch for set intersection
\item[EBD]: Edge Iterator with BinarySearch for set intersection (direction-oriented)
\item[EP]: Edge Iterator with Partitioning for set intersection
\item[EPD]: Edge Iterator with Partitioning for set intersection (direction-oriented)
\item[EH]: Edge Iterator with Hashing for set intersection
\item[EHD]: Edge Iterator with Hashing for set intersection (direction-oriented)
\item[F]: Forward
\item[FH]: Forward with Hashing
\item[FHD]: Forward with Hashing and degree-ordering
\item[TS]: Tri\_simple (Davis \cite{davis2018HPEC})
\item[LA]: Linear Algebra (CMU \cite{Low2017})
\item[CE]: Cover Edge (Bader, \cite{2023-BLGGLRD})
\item[CED]: Cover Edge with degree-ordering (Bader, \cite{2023-BLGGLRD})
\item[Bader]: this paper
\item[BaderD]: this paper with degree-ordering
\end{description}

\vspace{12pt}

While all of the algorithms tested have the same asymptotic worst-case complexity, the running times range by orders of magnitude between the approaches. In nearly every case where edge direction-orientation is used, the performance is typically improved by a constant factor up to two. The vertex-iterator and Itah-Rodeh algorithms are the slowest across the real and synthetic datasets. The timings between the Intel Xeon Platinum 8480+ and Intel Xeon Max 9480 are consistent, with the 8480+ a few percent faster than the 9480 processor. This is likely due to the fact that we are using single-threaded code on one core, and that the 8480+ is clocked at a slightly higher rate (2.00GHz vs 1.90GHz). 

In general, the forward algorithms and its variants tend to perform the fastest, followed by the edge-iterator, and then the vertex-iterator methods. The new fast triangle counting algorithm is competitive with the forward approaches, and may be useful when the results of a BFS are already available from the analyst's workflow, which is often the case. 

The performance of the road network graphs (roadNet-CA, roadNet-PA, roadNet-TX) are outliers from the other graphs. Road networks, unlike social networks, often have only low degree vertices (for instance, many degree four vertices), and large diameters. The percentage of horizontal edges ($k$) of these road networks is under 15\% and we see less benefit of the new approach due to this low value of $k$. In addition, the sorting of vertices by degree for the road network significantly harms the performance compared with the default ordering of the input. This may be due to the fact that there are few unique degree values, and sorting decimates the locality in the graph data structure.

The linear algebra approach \cite{Low2017} does not typically perform as well on the real and synthetic social networks. For example, on a large RMAT graph of scale 18, the linear algebra algorithm method takes seconds, whereas the new algorithm runs in under a second.  However, the linear algebra approach performs well on the road networks. 

\vspace{-0.05in}
\section{Conclusions}
\vspace{-0.05in}
\looseness=-1
In this paper we design and implement a novel, fast triangle counting algorithm, that uses new techniques to improve the performance. It is the first algorithm in decades to shine new light on triangle counting, and use a wholly new method of cover-edges to reduce the work of set intersections, rather than other approaches that are variants of the well-known vertex-iterator and edge-iterator methods.  We provide extensive performance results in a parsimonious framework for benchmarking serial triangle counting algorithms for sparse graphs in a uniform manner. The results use one of Intel's latest processor families, the Intel Sapphire Rapids (Platinum 8480+) and Sapphire Rapids HBM (CPU Max 9480) launched in the 1st quarter of 2023.  The new triangle counting algorithm can benefit when the results of a BFS are available, which is often the case in network science.  Additionally, this work will inspire much interest within the Graph Challenge community to implement versions of the presented algorithms for large-shared memory, distributed memory, GPU, or multi-GPU frameworks. 

\vspace{-0.05in}
\section{Future Work}
\vspace{-0.05in}
The fast triangle counting algorithm (Alg.~\ref{alg:TC}) can be readily parallelized using a parallel BFS, partitioning the edge set in parallel, and using a parallel triangle counting algorithm on graph $G0$, and parallelizing the set intersections for graph $G1$.  In future work, we will implement this parallel algorithm and compare its performance with other parallel approaches.

\vspace{-0.05in}
\section{Reproducibility}
\label{sec:reproducibility}
\vspace{-0.05in}
The sequential triangle counting source code is open source and available on GitHub at \url{https://github.com/Bader-Research/triangle-counting}.  The input graphs are from the Stanford Network Analysis Project (SNAP) available from \url{http://snap.stanford.edu/}.

\bibliographystyle{IEEEtran}
\bibliography{ref}

\onecolumn

\begin{sidewaystable}
\caption{Execution time (in seconds) for Intel Xeon 8480. \\ Key: IR: Itai-Rodeh. V: Vertex-iterator. VD: Vertex Iterator (direction-oriented). EM: Edge Iterator with MergePath. EMD: Edge Iterator with MergePath (direction-oriented). EB: Edge Iterator with BinarySearch. EBD: Edge Iterator with BinarySearch (direction-oriented). EP: Edge Iterator with Partitioning. EPD: Edge Iterator with Partitioning (direction-oriented). EH: Edge Iterator with Hashing. EHD: Edge Iterator with Hashing (direction-oriented). F: Forward. FH: Forward with Hashing. FHD: Forward with Hashing (degree-order). TS: Tri\_simple (Davis) LA: Linear Algebra (CMU). CE: Cover Edge. CED: Cover Edge (degree-order). Bader: this paper. BaderD: this paper with degree-order.}
\tiny
\centering
\begin{tabular}{lrrrrrrrrrrrrr}
 Graph          & n         & m         & \# triangles & IR         & V             & VD            & EM        & EMD       & EB        & EBD       & & $k$ (\%) \\ \hline
karate          & 34        & 78        & 45        & 0.000080      & 0.000019      & 0.000007      & 0.000012  & 0.000006  & 0.000018  & 0.000009  & & 35.9 \\
RMAT 6          & 64        & 1024      & 9100      & 0.000599      & 0.001716      & 0.000329      & 0.000404  & 0.000201  & 0.000828  & 0.000392  & & 93.8 \\
RMAT 7          & 128       & 2048      & 18855     & 0.003449      & 0.005494      & 0.001569      & 0.002001  & 0.001014  & 0.004508  & 0.002206  & & 90.9 \\
RMAT 8          & 256       & 4096      & 39602     & 0.005339      & 0.009073      & 0.002575      & 0.002865  & 0.001447  & 0.006661  & 0.003290  & & 87.6 \\
RMAT 9          & 512       & 8192      & 86470     & 0.017020      & 0.054048      & 0.014320      & 0.014765  & 0.007357  & 0.030639  & 0.015200  & & 87.2 \\
RMAT 10         & 1024      & 16384     & 187855    & 0.054833      & 0.093701      & 0.023941      & 0.020353  & 0.010080  & 0.045417  & 0.022546  & & 82.8 \\
RMAT 11         & 2048      & 32768     & 408876    & 0.168952      & 0.314510      & 0.077261      & 0.053464  & 0.026608  & 0.107207  & 0.053023  & & 81.1 \\
RMAT 12         & 4096      & 65536     & 896224    & 0.521454      & 1.083800      & 0.266666      & 0.140524  & 0.069853  & 0.287587  & 0.142856  & & 77.5 \\
RMAT 13         & 8192      & 131072    & 1988410   & 1.703786      & 3.735357      & 0.881334      & 0.372492  & 0.185039  & 0.725695  & 0.360435  & & 74.9 \\
RMAT 14         & 16384     & 262144    & 4355418   & 5.503025      & 13.078101     & 3.134624      & 0.982963  & 0.488870  & 1.792365  & 0.889259  & & 70.5 \\
RMAT 15         & 32768     & 524288    & 9576800   & 18.387938     & 45.361382     & 10.564304     & 2.615737  & 1.299128  & 4.741772  & 2.339147  & & 68.4 \\
RMAT 16         & 65536     & 1048576   & 21133772  & 60.224978     & 157.221897    & 35.818787     & 6.911778  & 3.434474  & 11.763857 & 5.776435  & & 65.5 \\
RMAT 17         & 131072    & 2097152   & 46439638  & 200.686691    & 549.218245    & 124.670216    & 18.305755 & 9.104815  & 30.948553 & 15.177363 & & 62.8 \\
RMAT 18         & 262144    & 4194304   & 101930789 & 665.063581    & 1890.988599   & 421.402839    & 48.163023 & 23.976539 & 78.511757 & 38.472993 & & 60.3 \\
amazon0302      & 262111    & 899792    & 717719    & 0.328209      & 0.239097      & 0.054081      & 0.137364  & 0.064924  & 0.190950  & 0.090092  & & 44.2 \\
amazon0312      & 400727    & 2349869   & 3686467   & 2.101546      & 1.331771      & 0.405438      & 0.714218  & 0.348752  & 1.003573  & 0.471135  & & 52.4 \\
amazon0505      & 410236    & 2439437   & 3951063   & 1.921873      & 1.541379      & 0.443954      & 0.753439  & 0.367458  & 1.069800  & 0.502501  & & 52.7 \\
amazon0601      & 403394    & 2443408   & 3986507   & 1.837775      & 1.445394      & 0.445080      & 0.750829  & 0.366805  & 1.072656  & 0.504409  & & 52.8 \\
loc-Brightkite  & 58228     & 214078    & 494728    & 0.538046      & 0.482041      & 0.135135      & 0.115865  & 0.057923  & 0.176090  & 0.087252  & & 43.2 \\
loc-Gowalla     & 196591    & 950327    & 2273138   & 5.330385      & 11.300779     & 3.612550      & 2.082631  & 1.035008  & 1.227245  & 0.593189  & & 50.8 \\
roadNet-CA      & 1971281   & 2766607   & 120676    & 0.588213      & 0.070164      & 0.032603      & 0.095966  & 0.061715  & 0.102809  & 0.067910  & & 14.5 \\
roadNet-PA      & 1090920   & 1541898   & 67150     & 0.329371      & 0.078338      & 0.036759      & 0.103484  & 0.066274  & 0.098181  & 0.038547  & & 14.6 \\
roadNet-TX      & 1393383   & 1921660   & 82869     & 0.454882      & 0.088063      & 0.022855      & 0.066163  & 0.042529  & 0.070018  & 0.046474  & & 14.0 \\
soc-Epinions1   & 75888     & 405740    & 1624481   & 4.561436      & 6.073439      & 1.503663      & 0.614835  & 0.304728  & 1.120885  & 0.552059  & & 53.3 \\
wiki-Vote       & 8297      & 100762    & 608389    & 0.496671      & 1.027744      & 0.175467      & 0.123803  & 0.061619  & 0.284746  & 0.141419  & & 54.3 \\[12pt]

Graph           & EP        & EPD       & EH        & EHD       & F         & FH        & FHD       & TS        & LA        & CE        & CED       & Bader     & BaderD \\ \hline
karate          & 0.000033  & 0.000015  & 0.000009  & 0.000005  & 0.000004  & 0.000004  & 0.000007  & 0.000005  & 0.000002  & 0.000006  & 0.000009  & 0.000009  & 0.000010 \\
RMAT 6          & 0.000928  & 0.000494  & 0.000064  & 0.000031  & 0.000033  & 0.000014  & 0.000016  & 0.000023  & 0.000023  & 0.000049  & 0.000042  & 0.000022  & 0.000021 \\
RMAT 7          & 0.005205  & 0.002781  & 0.000375  & 0.000170  & 0.000216  & 0.000068  & 0.000073  & 0.000120  & 0.000197  & 0.000456  & 0.000336  & 0.000125  & 0.000110 \\
RMAT 8          & 0.007306  & 0.003950  & 0.000467  & 0.000230  & 0.000334  & 0.000114  & 0.000113  & 0.000181  & 0.000309  & 0.000640  & 0.000510  & 0.000164  & 0.000177 \\
RMAT 9          & 0.036358  & 0.019781  & 0.002201  & 0.001108  & 0.001710  & 0.000548  & 0.000562  & 0.000910  & 0.001568  & 0.001564  & 0.001335  & 0.000409  & 0.000413 \\
RMAT 10         & 0.049769  & 0.027202  & 0.002840  & 0.001429  & 0.002386  & 0.000656  & 0.000668  & 0.001255  & 0.002187  & 0.003785  & 0.003294  & 0.000948  & 0.000937 \\
RMAT 11         & 0.129236  & 0.071272  & 0.007045  & 0.003576  & 0.006154  & 0.001528  & 0.001460  & 0.003294  & 0.005603  & 0.008990  & 0.007992  & 0.002005  & 0.002110 \\
RMAT 12         & 0.332512  & 0.184829  & 0.017642  & 0.008948  & 0.015684  & 0.003520  & 0.003302  & 0.008275  & 0.014181  & 0.021089  & 0.018934  & 0.004383  & 0.004499 \\
RMAT 13         & 0.852811  & 0.477536  & 0.044770  & 0.022457  & 0.040657  & 0.008140  & 0.007430  & 0.021401  & 0.035921  & 0.049489  & 0.044248  & 0.009687  & 0.009740 \\
RMAT 14         & 2.197651  & 1.241093  & 0.115374  & 0.056890  & 0.104338  & 0.018856  & 0.016352  & 0.056593  & 0.090133  & 0.113669  & 0.101336  & 0.020978  & 0.020060 \\
RMAT 15         & 5.611216  & 3.182339  & 0.316977  & 0.153824  & 0.271538  & 0.046304  & 0.038232  & 0.161525  & 0.228835  & 0.268428  & 0.236885  & 0.048083  & 0.043707 \\
RMAT 16         & 14.381678 & 8.204573  & 1.055079  & 0.514223  & 0.710508  & 0.119594  & 0.096148  & 0.502529  & 0.583421  & 0.634558  & 0.548067  & 0.118334  & 0.098932 \\
RMAT 17         & 37.158095 & 21.256965 & 3.256451  & 1.640081  & 1.864839  & 0.334746  & 0.251170  & 1.451738  & 1.484360  & 1.497880  & 1.261575  & 0.330477  & 0.249333 \\
RMAT 18         & 93.966516 & 54.116325 & 9.170543  & 4.523967  & 4.845280  & 0.909296  & 0.611022  & 3.957162  & 3.743856  & 3.509278  & 2.888053  & 0.907862  & 0.624522 \\
amazon0302      & 0.314972  & 0.209145  & 0.064610  & 0.032767  & 0.024226  & 0.020652  & 0.041335  & 0.037947  & 0.022963  & 0.044121  & 0.066253  & 0.043772  & 0.060910 \\
amazon0312      & 1.577489  & 0.888635  & 0.289698  & 0.137439  & 0.107685  & 0.075531  & 0.139509  & 0.164931  & 0.100935  & 0.147793  & 0.198154  & 0.129337  & 0.188718 \\
amazon0505      & 1.680979  & 0.942646  & 0.302644  & 0.143367  & 0.113673  & 0.078874  & 0.146358  & 0.172100  & 0.106964  & 0.154702  & 0.208559  & 0.130193  & 0.178938 \\
amazon0601      & 1.675168  & 0.941213  & 0.300678  & 0.142704  & 0.114721  & 0.080125  & 0.145550  & 0.170971  & 0.107216  & 0.155132  & 0.208443  & 0.130382  & 0.178437 \\
loc-Brightkite  & 0.262203  & 0.152707  & 0.026410  & 0.013830  & 0.011533  & 0.006663  & 0.010069  & 0.013557  & 0.011688  & 0.011810  & 0.015568  & 0.008635  & 0.013218 \\
loc-Gowalla     & 2.840775  & 2.060617  & 0.358680  & 0.175656  & 0.085123  & 0.039745  & 0.056709  & 0.188951  & 0.079884  & 0.074447  & 0.091355  & 0.046396  & 0.064264 \\
roadNet-CA      & 0.167447  & 0.093240  & 0.067984  & 0.049379  & 0.032793  & 0.038295  & 0.172404  & 0.051017  & 0.039746  & 0.081306  & 0.209616  & 0.132360  & 0.242484 \\
roadNet-PA      & 0.095960  & 0.053119  & 0.038335  & 0.027772  & 0.018091  & 0.021191  & 0.084368  & 0.028920  & 0.022348  & 0.042012  & 0.110022  & 0.128744  & 0.248665 \\
roadNet-TX      & 0.117129  & 0.065016  & 0.047048  & 0.034324  & 0.022388  & 0.026438  & 0.108113  & 0.035132  & 0.027658  & 0.052228  & 0.140249  & 0.154466  & 0.309297 \\
soc-Epinions1   & 1.538742  & 0.876433  & 0.100463  & 0.048663  & 0.062346  & 0.019989  & 0.023260  & 0.051619  & 0.063603  & 0.043226  & 0.042753  & 0.021945  & 0.023399 \\
wiki-Vote       & 0.354843  & 0.189848  & 0.020137  & 0.010243  & 0.018043  & 0.005117  & 0.005159  & 0.009498  & 0.019525  & 0.013211  & 0.014798  & 0.005232  & 0.005956 
\end{tabular}
\label{t:results:intel8480}
\end{sidewaystable}

\begin{sidewaystable}
\caption{Execution time (in seconds) for Intel Xeon Max 9480. \\ Key: IR: Itai-Rodeh. V: Vertex-iterator. VD: Vertex Iterator (direction-oriented). EM: Edge Iterator with MergePath. EMD: Edge Iterator with MergePath (direction-oriented). EB: Edge Iterator with BinarySearch. EBD: Edge Iterator with BinarySearch (direction-oriented). EP: Edge Iterator with Partitioning. EPD: Edge Iterator with Partitioning (direction-oriented). EH: Edge Iterator with Hashing. EHD: Edge Iterator with Hashing (direction-oriented). F: Forward. FH: Forward with Hashing. FHD: Forward with Hashing (degree-order). TS: Tri\_simple (Davis) LA: Linear Algebra (CMU). CE: Cover Edge. CED: Cover Edge (degree-order). Bader: this paper. BaderD: this paper with degree-order.}
\tiny
\centering
\begin{tabular}{lrrrrrrrrrrrrr}
 Graph          & n         & m         & \# triangles & IR         & V             & VD            & EM        & EMD       & EB        & EBD       & & $k$ (\%) \\ \hline
karate          & 34        & 78        & 45        & 0.000084 & 0.000019 & 0.000007 & 0.000012 & 0.000006 & 0.000018 & 0.000009       & & 35.9 \\	 
RMAT 6          & 64        & 1024      & 9100      & 0.000653 & 0.001432 & 0.000291 & 0.000392 & 0.000192 & 0.000869 & 0.000426       & & 93.8 \\	 
RMAT 7          & 128       & 2048      & 18855     & 0.003678 & 0.005776 & 0.001651 & 0.002121 & 0.001065 & 0.004745 & 0.002365       & & 90.9 \\	 
RMAT 8          & 256       & 4096      & 39602     & 0.005814 & 0.009776 & 0.002655 & 0.003127 & 0.001569 & 0.007266 & 0.003583       & & 87.6 \\	 
RMAT 9          & 512       & 8192      & 86470     & 0.018371 & 0.030974 & 0.008197 & 0.008432 & 0.004211 & 0.017575 & 0.008738       & & 87.2 \\	 
RMAT 10         & 1024      & 16384     & 187855    & 0.058950 & 0.172546 & 0.026058 & 0.021977 & 0.010949 & 0.049523 & 0.024602       & & 82.8 \\	 
RMAT 11         & 2048      & 32768     & 408876    & 0.183945 & 0.342650 & 0.084153 & 0.058176 & 0.028950 & 0.116835 & 0.057881       & & 81.1 \\	 
RMAT 12         & 4096      & 65536     & 896224    & 0.566386 & 1.180451 & 0.290440 & 0.152799 & 0.076057 & 0.313237 & 0.155666       & & 77.5 \\	 
RMAT 13         & 8192      & 131072    & 1988410   & 1.855761 & 4.068261 & 0.960160 & 0.404467 & 0.202029 & 0.789813 & 0.392430       & & 74.9 \\	 
RMAT 14         & 16384     & 262144    & 4355418   & 6.005944 & 14.231056 & 3.411291 & 1.069533 & 0.532122 & 1.951912 & 0.968742      & & 70.5 \\	 
RMAT 15         & 32768     & 524288    & 9576800   & 20.030438 & 49.387592 & 11.505111 & 2.839230 & 1.412183 & 5.141348 & 2.544929    & & 68.4 \\	 
RMAT 16         & 65536     & 1048576   & 21133772  & 65.629646 & 171.271757 & 38.996294 & 7.518746 & 3.732511 & 12.724100 & 6.271381  & & 65.5 \\	 
RMAT 17         & 131072    & 2097152   & 46439638  & 375.735104 & 485.577535 & 135.059792 & 19.910985 & 9.883762 & 33.311904 & 16.416130  & & 62.8 \\	 
RMAT 18         & 262144    & 4194304   & 101930789 & 1268.171131 & 1659.316098 & 457.361748 & 53.163797 & 26.201594 & 85.430737 & 41.790449  & & 60.3 \\	 
amazon0302      & 262111    & 899792    & 717719    & 0.338818 & 0.355492 & 0.056826 & 0.135808 & 0.066465 & 0.192788 & 0.093958       & & 44.2 \\	 
amazon0312      & 400727    & 2349869   & 3686467   & 2.144190 & 1.417495 & 0.430188 & 0.734345 & 0.362226 & 1.036567 & 0.494136       & & 52.4 \\	 
amazon0505      & 410236    & 2439437   & 3951063   & 1.953721 & 1.543616 & 0.471659 & 0.774813 & 0.382143 & 1.106683 & 0.528466       & & 52.7 \\	 
amazon0601      & 403394    & 2443408   & 3986507   & 1.864989 & 1.531052 & 0.468426 & 0.771804 & 0.381114 & 1.110423 & 0.626398       & & 52.8 \\	 
loc-Brightkite  & 58228     & 214078    & 494728    & 0.583671 & 0.603258 & 0.147395 & 0.125952 & 0.063082 & 0.191693 & 0.094928       & & 43.2 \\	 
loc-Gowalla     & 196591    & 950327    & 2273138   & 5.787065 & 12.271734 & 3.931624 & 2.254498 & 1.122959 & 1.317039 & 0.641545      & & 50.8 \\	 
roadNet-CA      & 1971281   & 2766607   & 120676    & 0.653579 & 0.078303 & 0.037398 & 0.106735 & 0.069202 & 0.114088 & 0.075968       & & 14.5 \\	 
roadNet-PA      & 1090920   & 1541898   & 67150     & 0.361385 & 0.044707 & 0.021273 & 0.060422 & 0.039120 & 0.064219 & 0.042327       & & 14.6 \\	 
roadNet-TX      & 1393383   & 1921660   & 82869     & 0.498325 & 0.052356 & 0.025527 & 0.072977 & 0.047387 & 0.077212 & 0.051569       & & 14.0 \\	 
soc-Epinions1   & 75888     & 405740    & 1624481   & 4.949491 & 6.610313 & 1.637538 & 0.670166 & 0.332638 & 1.226177 & 0.602162       & & 53.3 \\	 
wiki-Vote       & 8297      & 100762    & 608389    & 0.564122 & 1.101121 & 0.191629 & 0.134719 & 0.067070 & 0.309990 & 0.154091       & & 54.3 \\[12pt]

Graph           & EP        & EPD       & EH        & EHD       & F         & FH        & FHD       & TS        & LA        & CE        & CED       & Bader     & BaderD \\ \hline
karate          & 0.000034 & 0.000015 & 0.000009 & 0.000006 & 0.000004 & 0.000004 & 0.000008 & 0.000005 & 0.000002 & 0.000006 & 0.000009 & 0.000009 & 0.000011   \\	
RMAT 6          & 0.001014 & 0.000532 & 0.000071 & 0.000033 & 0.000034 & 0.000016 & 0.000017 & 0.000025 & 0.000027 & 0.000057 & 0.000046 & 0.000023 & 0.000023	 \\	
RMAT 7          & 0.005489 & 0.002928 & 0.000398 & 0.000177 & 0.000230 & 0.000073 & 0.000081 & 0.000131 & 0.000204 & 0.000457 & 0.000358 & 0.000133 & 0.000120	 \\	
RMAT 8          & 0.007966 & 0.004306 & 0.000511 & 0.000249 & 0.000367 & 0.000127 & 0.000126 & 0.000194 & 0.000338 & 0.000675 & 0.000551 & 0.000187 & 0.000199	 \\	
RMAT 9          & 0.020866 & 0.011328 & 0.001285 & 0.000644 & 0.000981 & 0.000316 & 0.000318 & 0.000538 & 0.000905 & 0.001720 & 0.001442 & 0.000463 & 0.000477 	 \\	
RMAT 10         & 0.054240 & 0.029602 & 0.003113 & 0.001598 & 0.002603 & 0.000729 & 0.000703 & 0.001370 & 0.002384 & 0.004150 & 0.003601 & 0.001097 & 0.001032 	 \\	
RMAT 11         & 0.140766 & 0.077608 & 0.007638 & 0.003886 & 0.006698 & 0.001663 & 0.001579 & 0.003562 & 0.006106 & 0.009849 & 0.008694 & 0.002186 & 0.002306 	 \\	
RMAT 12         & 0.361722 & 0.201107 & 0.019090 & 0.009670 & 0.017082 & 0.003810 & 0.003632 & 0.008985 & 0.015445 & 0.023041 & 0.020660 & 0.004851 & 0.004978	 \\	
RMAT 13         & 0.927790 & 0.519227 & 0.048198 & 0.024338 & 0.044297 & 0.008913 & 0.008062 & 0.022965 & 0.039156 & 0.054002 & 0.048318 & 0.010538 & 0.010336	 \\	
RMAT 14         & 2.390258 & 1.350289 & 0.123571 & 0.061666 & 0.113511 & 0.020460 & 0.017647 & 0.059840 & 0.098365 & 0.124021 & 0.110538 & 0.023147 & 0.021745	 \\	
RMAT 15         & 6.181050 & 3.463447 & 0.335935 & 0.164295 & 0.294636 & 0.049240 & 0.040570 & 0.167853 & 0.247720 & 0.291674 & 0.256692 & 0.052713 & 0.047445	 \\	
RMAT 16         & 15.620494 & 8.921255 & 1.139298 & 0.552865 & 0.766895 & 0.123122 & 0.098112 & 0.528412 & 0.628183 & 0.687476 & 0.591732 & 0.127122 & 0.107139	 \\	
RMAT 17         & 40.535918 & 23.082879 & 3.430538 & 1.699966 & 1.983156 & 0.334009 & 0.243371 & 1.475421 & 1.577658 & 1.602727 & 1.358685 & 0.336016 & 0.249839  \\	
RMAT 18         & 103.623609 & 58.983807 & 10.785311 & 5.093805 & 5.185930 & 0.929694 & 0.605976 & 5.027324 & 4.067794 & 3.904872 & 3.126732 & 0.936819 & 0.629180   \\	
amazon0302      & 0.301605 & 0.157618 & 0.064183 & 0.033356 & 0.024741 & 0.021114 & 0.042219 & 0.038521 & 0.023865 & 0.045608 & 0.066787 & 0.046098 & 0.065133	 \\	
amazon0312      & 1.671686 & 0.947525 & 0.285707 & 0.139874 & 0.106593 & 0.075217 & 0.143323 & 0.159351 & 0.103863 & 0.152911 & 0.207114 & 0.137305 & 0.185652	 \\	
amazon0505      & 1.780392 & 1.006270 & 0.296284 & 0.145469 & 0.112280 & 0.078005 & 0.150319 & 0.166425 & 0.109580 & 0.159629 & 0.217396 & 0.141835 & 0.192150	 \\	
amazon0601      & 1.807190 & 1.005138 & 0.311293 & 0.145677 & 0.113539 & 0.079493 & 0.148641 & 0.176862 & 0.109842 & 0.165447 & 0.217130 & 0.144488 & 0.192724	 \\	
loc-Brightkite  & 0.285579 & 0.166176 & 0.028593 & 0.015042 & 0.012540 & 0.007266 & 0.010768 & 0.014710 & 0.012744 & 0.012909 & 0.016484 & 0.009446 & 0.013962	 \\	
loc-Gowalla     & 3.057021 & 2.229423 & 0.377311 & 0.187272 & 0.088885 & 0.040840 & 0.056879 & 0.195321 & 0.084664 & 0.079257 & 0.093252 & 0.049751 & 0.069627	 \\	
roadNet-CA      & 0.184792 & 0.103255 & 0.076952 & 0.056482 & 0.037726 & 0.044601 & 0.191125 & 0.059184 & 0.045363 & 0.095519 & 0.231007 & 0.141794 & 0.262101	 \\	
roadNet-PA      & 0.104778 & 0.058080 & 0.042658 & 0.031094 & 0.020907 & 0.024783 & 0.094690 & 0.032229 & 0.024773 & 0.048739 & 0.121119 & 0.070754 & 0.141447	 \\	
roadNet-TX      & 0.127673 & 0.071134 & 0.052878 & 0.039065 & 0.025792 & 0.030903 & 0.121140 & 0.040004 & 0.031081 & 0.062052 & 0.154108 & 0.088572 & 0.180140	 \\	
soc-Epinions1   & 1.674840 & 0.955909 & 0.110867 & 0.054109 & 0.067796 & 0.021483 & 0.024809 & 0.058189 & 0.070127 & 0.047488 & 0.045225 & 0.023882 & 0.025212	 \\	
wiki-Vote       & 0.387609 & 0.207687 & 0.021763 & 0.011102 & 0.019645 & 0.005587 & 0.005645 & 0.010345 & 0.021255 & 0.014397 & 0.015981 & 0.005747 & 0.006269	 
\end{tabular}
\label{t:results:intel9480}
\end{sidewaystable}


\end{document}